\documentclass[prl,twocolumn,aps,amssymb,amsmath,tightenlines]{revtex4}
\usepackage{graphicx,epsfig}

\begin{document}

\title{Two-dimensional wave propagation without anomalous dispersion}
\author{Carl M. Bender$^{a,b}$}\email{cmb@wustl.edu}
\author{Francisco J. Rodr\'iguez Fort\~uno$^b$}\email{francisco.rodriguez_fortuno@kcl.ac.uk}
\author{Sarben Sarkar$^b$}\email{sarben.sarkar@kcl.ac.uk}
\author{Anatoly V. Zayats$^b$}\email{a.zayats@kcl.ac.uk}

\affiliation{$^a$Department of Physics, Washington University, St. Louis, MO
63130, USA\\
$^b$Department of Physics, King's~College~London, London WC2R 2LS, UK}

\begin{abstract}
In two space dimensions and one time dimension a wave changes its shape even in
the absence of a dispersive medium. However, this anomalous dispersive behavior
in empty two-dimensional space does not occur if the wave dynamics is described
by a linear homogeneous wave equation in two space dimensions and {\it two} time
dimensions. Wave propagation in such a space can be realized in a
three-dimensional anisotropic metamaterial in which one of the space dimensions
has a negative permittivity and thus serves as an effective second time
dimension. These results lead to a fundamental understanding and new approaches
to ultrashort pulse shaping in nanostructures and metamaterials.
\end{abstract}

\date{\today}
\maketitle

A light pulse traveling through a 3-D medium (D is the number of space
dimensions) changes its shape and develops a tail as it propagates if different
frequency components of the wave travel at different speeds. This pulse
broadening, which is called {\it dispersion}, occurs in a medium with a
frequency-dependent refractive index. Dispersion of light does not occur in 
3-D in a vacuum.

In contrast, in 2-D (three-dimensional space-time) a pulse changes its shape and
forms a tail in a {\it vacuum} even though the wave speed $c$ is a
frequency-independent constant. This dispersion in vacuum in 2-D cannot be
explained in terms of a variable wave speed, and so this phenomenon is called
{\it anomalous dispersion} \cite{R1,R2}. Anomalous dispersion in vacuum is a
geometrical phenomenon associated with the dimension of space in which the waves
propagate and it occurs for all kinds of waves, electromagnetic and acoustic.
It occurs in two-dimensional but not in three-dimensional space.

The rumbling of thunder provides physical evidence of anomalous dispersion. A
lightning bolt is a long (approximately translationally invariant) vertical line
source. The sound waves that are produced propagate outward as {\it
two}-dimensional cylindrical waves. Although the lightning bolt is
instantaneous, an observer does not hear a ``bang,'' but rather a rumbling that
fades like $1/t$.

In this Letter we show that by adding a second time-like coordinate to a
three-dimensional space-time, anomalous dispersion in vacuum can be eliminated,
leading to nondispersing pulse propagation. Such a non-Minkowskian space-time
can be simulated by metamaterials with hyperbolic dispersion \cite{R3}. These
metamaterials have two space directions in which their effective permittivity is
positive, and a third direction with a negative permittivity. Electromagnetic
fields in a nondispersive and nonmagnetic anisotropic material with a uniaxial
dielectric permittivity tensor $\overline{\overline\varepsilon}=\mathrm{diag}(
\varepsilon^x,\varepsilon^x,\varepsilon^z)$ obey Maxwell's equations $\nabla
\times\mathbf{E}=-\mu_0\mathbf{H}_t$ and $\nabla\times\mathbf{H}=\varepsilon_0
\overline{\overline\varepsilon}\mathbf{E}_t$. These can be combined to give
$\nabla\times\nabla\times\mathbf{E}=-\mu_0\varepsilon_0\overline{\overline
\varepsilon}\mathbf{E}_{tt}$. This, together with Gauss' law $\nabla\cdot\left(
\overline{\overline\varepsilon}\mathbf{E}\right)=0$, gives the wave equation
\cite{R3}
$$u_{tt}=\frac{c^2}{\varepsilon^x}\left(u_{xx}+u_{yy}\right)+\frac{c^2}{
\varepsilon^z}u_{zz},$$
where the scalar field $u\equiv E^z$ represents the $z$ component of the
electric field and the speed of light is given by $c^2=(\mu_0\varepsilon_0)^{-1
}$. This equation becomes the usual homogeneous 3-D 
wave equation when $\varepsilon^x=\varepsilon^z$, as is the case in vacuum.
However, it can also represent the wave equation in two time and two space
dimensions when $\varepsilon^z$ is negative and $\varepsilon^x$ is positive.
Such materials exist and are known as {\it hyperbolic materials}; anisotropic
composites with diagonal components of the effective permittivity tensor having
opposite signs can be constructed as metal-dielectric multilayers or metal
nanorod arrays \cite{R4}. Light propagates inside them as cones \cite{R5,R6} and
their two-time character has been studied \cite{R3,R7,R8}, raising the
intriguing possibility of observing dispersionless propagation in 2-D.
Space-times with two time dimensions also occur in M-theory \cite{R9}.

The homogeneous linear wave equation
\begin{equation}
u_{tt}=c^2\nabla^2u
\label{E1}
\end{equation}
describes waves $u({\bf x},t)$ that travel with constant (frequency-independent)
wave speed $c$ through a uniform medium. This wave equation describes how an
initial pulse at $t=0$ given by the initial conditions
\begin{equation}
u({\bf x},0)=q({\bf x}),\quad u_t({\bf x},0)=p({\bf x}),
\label{E2}
\end{equation}
evolves into the wave $u({\bf x},t)$ at time $t$. This initial-value problem for
the wave equation has an explicit quadrature solution in any space dimension.

\begin{widetext}

\begin{figure}[h!]
\begin{center}
\includegraphics[trim=1mm 0mm 0mm 0mm,clip=true,scale=0.90]{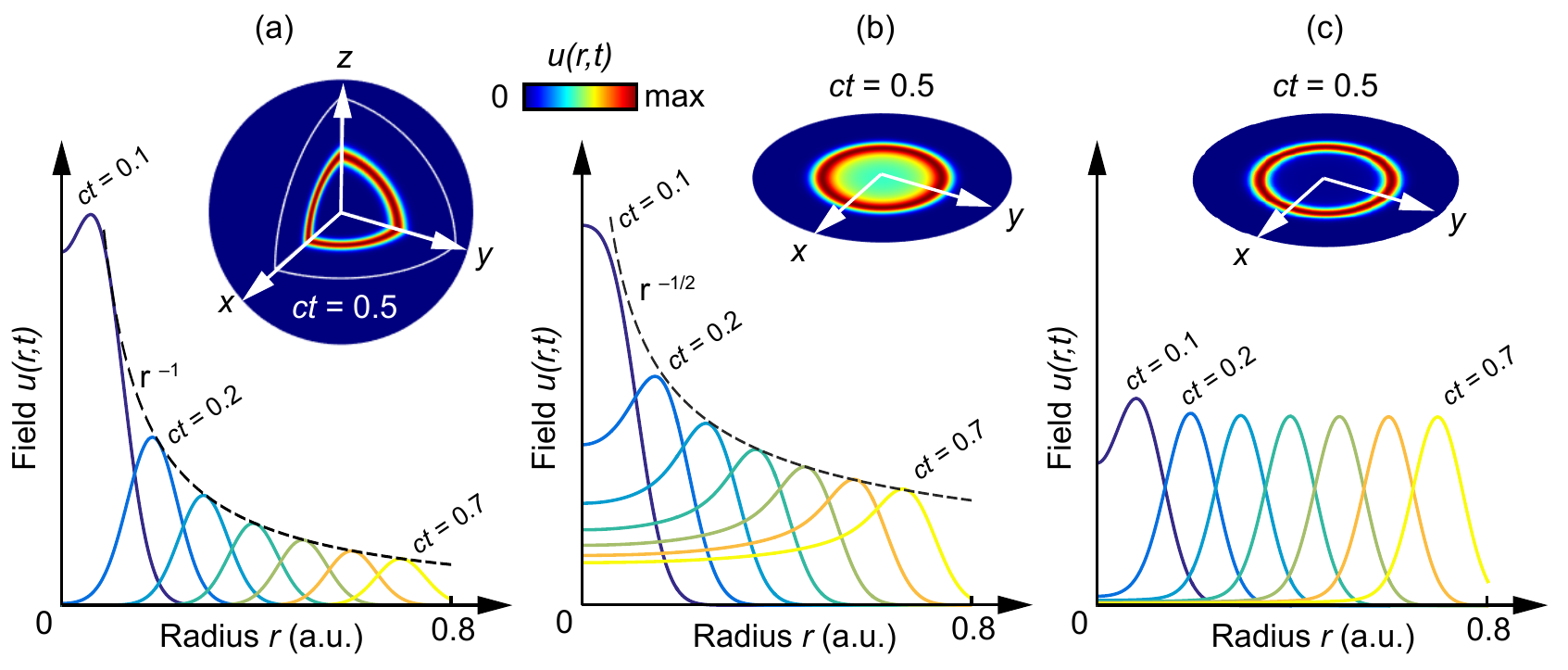}
\end{center}
\caption{[Color online] Numerical solution of the wave equation in three cases:
(a) conventional 3-D wave equation $u_{tt}=c^2(u_{xx}+u_{yy}+u{zz})$; (b)
conventional 2-D wave equation $u_{tt}=c^2(u_{xx}+u_{yy})$ showing anomalous
dispersion; (c) Modified two-dimensional wave equation $u_{tt}-(1/t)u_t+(1/t^2)
u=c^2(u_{xx}+u_{yy})$ with no anomalous dispersion. The insets show the solution
at a fixed time. The initial conditions are: zero initial value $u(r,t=0)=0$ and
a radially symmetric initial time derivative gaussian distribution $u_{t}(r,t=0)
= exp(-r^2/(2\sigma_r^2))$ with $\sigma_r=0.01$. The radial variable is $r=
\sqrt{x^2+y^2+z^2}$ for (a) and $r=\sqrt{x^2+y^2}$ for (b) and (c).}
\label{F1}
\end{figure}

\end{widetext}

Wave propagation in odd-dimensional space is fundamentally different from
wave propagation in even-dimensional space. When the space dimension D is odd
and ${\rm D}>1$, waves obey {\it Huygens' principle} \cite{R1,R2}; that is,
waves created by an instantaneous point source at $t=0$ ({\it e.g.}, a light
pulse) take the form of an expanding bubble. After the wavefront passes by, the
medium instantly returns to quiescence. An observer sees blackness until the
wave arrives, sees an instantaneous flash as the wave passes by, and immediately
afterward sees blackness again [see Fig.~\ref{F1}(a)]. Such a wave propagates on
the surface of the light cone. In contrast, in even-dimensional space an
instantaneous point source gives rise to a wave that develops a tail. An
observer sees blackness until the wave arrives and then sees a flash. However,
the medium does not immediately return to quiescence; rather, the wave amplitude
decays to 0 like $t^{-\alpha}$, where $\alpha>0$ depends on D. When ${\rm D}=2$,
the wave amplitude decays to 0 like $1/t$. The tail of the wave propagates less
rapidly than $c$ as a consequence of anomalous dispersion. Such a wave
propagates on the surface {\it and in the interior} of the light cone [see
Fig.~\ref{F1}(b)]. In a vacuum, if Huygens' principle applies, there is no
anomalous dispersion \cite{R1,R2}.

{\bf 2-D wave propagation without anomalous dispersion.} Anomalous dispersion
occurs in vacuum in 2-D. However, one might wonder whether it is possible to
produce a medium whose dispersive properties exactly cancel the anomalous
dispersion that occurs in 2-D wave propagation. Such a medium actually exists.
The effect of anomalous dispersion can be cancelled if we modify the 2-D wave
equation by adding lower derivatives in time:
\begin{equation}
u_{tt}-t^{-1}u_t+t^{-2}u=c^2(u_{xx}+u_{yy}).
\label{E3}
\end{equation}
The one-derivative term models gain, which speeds up the lagging tail of a
2-D wave. However, this term is too strong, so to reduce its effect we also
introduce the time-dependent term $t^{-2}u$. For this artificial medium a flash
bulb gives rise to a wave that does not disperse; the wave remains confined to
the surface of the light cone and does not leak into the interior of the light
cone [see Fig.~\ref{F1}(c)]. Below we demonstrate analytically this
dispersionless propagation. Furthermore, we show that the wave equation
(\ref{E3}) is equivalent to a {\it constant-coefficient} wave equation in {\it
two} space and {\it two} time dimensions.

{\bf 1-D homogeneous wave equation.} The 1-D wave equation is special because
its solutions obey Huygens' principle when $p=0$, $q\neq0$, and they exhibit
anomalous dispersion when $q=0$, $p\neq0$. The general solution $u(x,t)=
f(x+ct)+g(x-ct)$ to the 1-D wave equation is a superposition of two waves of
unchanging shape traveling at constant speed $c$, one moving to the left and the
other moving to the right. The exact solution satisfying the initial conditions
(\ref{E2}) is given by {\it D'Alembert's formula}:
\begin{equation}
u(x,t)=\frac{q(x+ct)+q(x-ct)}{2}+{1\over 2c}\int_{x-ct}^{x+ct}ds\,p(s).
\label{E4}
\end{equation}

{\bf 3-D homogeneous wave equation.} The exact solution $u({\bf x},t)$ to the
initial-value problem for the 3-D wave equation $u_{tt}=c^2(u_{xx}+u_{yy}+u_{zz}
)$ is obtained by using a construction invented by Kirchhoff. The quadrature
solution is given compactly by {\it Poisson's formula} \cite{R1,R2}:
\begin{equation}
u(x,y,z,t)={\partial\over\partial t}(t\omega_{ct}[q])+t\omega_{ct}[p].
\label{E5}
\end{equation}
The {\it spherical mean} $\omega_{ct}[\phi]$ of the function $\phi(x,y,z)$ is an
integral over the surface of a three-dimensional sphere of radius $ct$ centered
at $(x,y,z)$:
\begin{equation}
\omega_{ct}[\phi]\equiv\int_{\alpha^2+\beta^2+\gamma^2=1}{d\Omega\over4\pi}\>
\phi(x+ct\alpha,\>y+ct\beta,\>z+ct\gamma),
\label{E6}
\end{equation}
where $d\Omega$ is an infinitesimal solid angle. Equations (\ref{E5}) and
(\ref{E6}) are the complete solution to (\ref{E1}) and (\ref{E2}).

{\bf 2-D homogeneous wave equation.} The solution to the initial-value problem
for the 2-D wave equation $u_{tt}=c^2(u_{xx}+u_{yy})$ can be expressed in
Poisson form (\ref{E5}), but now the 3-D spherical mean is a weighted average
over the surface of a 2-D disk centered at $(x,y)$:
\begin{equation}
\omega_{ct}[\phi]\equiv{1\over2\pi}\int\int_{\alpha^2+\beta^2\leq1}d\alpha\,
d\beta\>{\phi(x+ct\alpha,\>y+ct\beta)\over\sqrt{1-\alpha^2-\beta^2}}.
\label{E7}
\end{equation}
We derive the integral in (\ref{E7}) from (\ref{E6}) by applying Hadamard's {\it
method of descent} in which we project from ${\rm D}=3$ down to ${\rm D}=2$ by
assuming that $\phi(x,y,z)$ is independent of $z$.

The method of descent may be used to project the 2-D solution (\ref{E5}) and
(\ref{E7}) down to ${\rm D}=1$, allowing us to recover D'Alembert's solution
(\ref{E4}) to the 1-D wave equation. To do so we choose $\phi(x,y)$ in
(\ref{E7}) to be independent of $y$.

{\bf Verification of Huygens' Principle in 3-D.} The 3-D solution $u(x,y,z,t)$
in (\ref{E5})-(\ref{E6}) depends on values of $q$ and $p$ only at points that
are {\it exactly} a distance $ct$ from $(x,y,z)$. Points that are further from
$(x,y,z)$ than $ct$ do not affect the solution $u(x,y,z,t)$ because wave
disturbances from such points cannot travel faster than $c$. However, the fact
that points that are {\it closer} to $(x,y,z)$ than $ct$ also do not affect the
solution $u(x,y,z,t)$ is a surprise because there is ample time for waves
emanating from these nearby points to reach the point $(x,y,z)$. It is this
feature of 3-D wave propagation that leads to Huygens' principle. As stated
earlier, 1-D and 2-D wave propagation do not obey Huygens' principle.


{\bf Solution to the initial-value problem for a 3-D point disturbance.}
Consider a 3-D medium that is initially quiescent and suppose that at $t=0$
there is a momentary light pulse at the origin. We represent such a disturbance
by
\begin{equation}
u(x,y,z,0)=0,\quad u_t(x,y,z,0)=\delta(x)\delta(y)\delta(z).
\label{E8}
\end{equation}
To see how this disturbance propagates in time, we substitute (\ref{E8}) into
Poisson's formula (\ref{E5}) and evaluate the integral in (\ref{E6}). We obtain
the spherical wave form
\begin{equation}
u(x,y,z,t)=\textstyle{\frac{1}{4\pi cr}}\delta(r-ct),
\label{E9}
\end{equation}
where $r=\sqrt{x^2+y^2+z^2}$. This 3-D wave resulting from the point disturbance
(\ref{E8}) is precisely the expected expanding bubble. An observer at a distance
$r$ from the 3-D point disturbance waits a time $t=r/c$ and then detects a
momentary flash followed by {\it total} quiescence. There is no remnant of this
disturbance when $t>r/c$.

{\bf Anomalous dispersion in 1-D.} Huygens' principle does not hold in 1-D
because D'Alembert's solution (\ref{E4}) for $u(x,t)$ depends on $p(s)$ for $x-c
t\leq s\leq x+ct$ and not just on $p(x+ct)$ and $p(x-ct)$. To see this, consider
the evolution of a 1-D point disturbance at $x=0$ (a pulse), which we
represent by the initial conditions $q(x)=u(x,0)=0$, $p(x)=u_t(x,0)=\delta(x)$.
D'Alembert's formula (\ref{E4}) shows that these initial conditions spawn a wave
in the form of a two-sided step function:
$$u(x,t)=\textstyle{{1\over 2c}}\>\theta(ct-|x|).$$
An observer at $x$ must wait a time $t=|x|/c$ before the pulse arrives. After
the wavefront passes, the medium does not return to its initially quiescent
state; an upward displacement of $1/(2c)$ {\it persists for all time.} Thus, 1-D
wave propagation violates Huygens' principle \cite{R1,R2}.

There is a special class of initial conditions, $p(x)=u_t(x,0)=0$, that creates
waves that {\it do} obey Huygens' principle. This is because waves arising
solely from an initial displacement leave the medium quiescent after they have
passed. For example, consider the initial conditions
$$u_t(x,0)=0,\quad u(x,0)=\left\{\begin{array}{cl}1-|x| & (|x|<1),\\ 0 &
(|x|\geq 1).\end{array}\right.$$
These initial conditions correspond to an initially triangular transverse
displacement and no initial transverse velocity. D'Alembert's formula shows that
the initial pulse splits into a right-going and a left-going triangular pulse,
each having half the initial amplitude. The two pulses travel to the right and
to the left with speed $c$ and do not change their shape. The medium is
quiescent until a pulse arrives, and after the pulse passes by, the medium
returns to quiescence.

{\bf 2-D waves and anomalous dispersion.} To show that 2-D wave propagation does
not obey Huygens' principle, we examine the effect of a 2-D point disturbance:
\begin{equation}
u(x,y,0)=0,\quad u_t(x,y,0)=\delta(x)\delta(y).
\label{E10}
\end{equation}
This pulse disturbance is an idealization of the initial conditions created,
for example, by a fish surfacing on a quiet pond. Substituting (\ref{E10}) into
(\ref{E7}) and (\ref{E5}), we obtain
\begin{eqnarray}
u(x,y,t)=\frac{\theta(c^2t^2-x^2-y^2)}{2\pi c\sqrt{c^2t^2-x^2-y^2}}.
\label{E11}
\end{eqnarray}
This formula describes a circular wave that propagates outward at speed $c$. An
observer at $x^2+y^2=r^2$ must wait until time $t=r/c$ before the leading edge
of the wave front arrives. The value of $u$ at the leading edge is infinite, but
after the wavefront passes, the trailing wave decays like $1/t$ for large $t$.

This explains why thunder rumbles, as mentioned previously.
The surface of the earth is locally flat and
two-dimensional. The sound of thunder propagates across the surface in the same
way that the wave in (\ref{E11}) propagates from a pointlike disturbance. A less
idealized way to understand the rumbling of thunder is to view it as a 3-D wave
in which the observer on the ground hears waves of the form (\ref{E8}) that
emanate from successively higher and more distant points on the lightning bolt
as $t$ increases. This explains the continued and tapering rumbling. Also,
since the height of the lightning bolt is finite, after some time the rumbling
abruptly stops.

{\bf Large-time asymptotic behavior of 2-D waves.} The $1/t$ decay in 2-D space
of a trailing wave, as in (\ref{E11}), is a general property of any wave created
by a localized disturbance. To show this, substitute $a=ct\alpha$ and $b=ct
\beta$ into (\ref{E7}). Assuming that $\phi$ has compact support, then
\begin{equation}
t\omega_{ct}[\phi]\sim{1\over2\pi c^2t}{\int\int}_R da\,db\,\phi(a,b)\quad(t\to
\infty).
\label{E12}
\end{equation}
If the integral in (\ref{E12}) exists and is nonzero, then $t\omega_{ct}[\phi]
\propto1/t$ as $t\to\infty$.

{\bf Solution to the modified wave equation (\ref{E3}).} This wave equation is
singular at $t=0$, so we do not try to solve (\ref{E3}) for the general initial
conditions (\ref{E2}). However, we impose the special initial conditions
$$u(x,y,0)=q(x,y)=0,\quad u_t(x,y,0)=p(x,y)$$
by following the Kirchhoff construction procedure. We rewrite (\ref{E3}) in
spherical coordinates and seek radially symmetric solutions $u(x,t)$ to
\begin{equation}
u_{tt}-t^{-1}u_t+t^{-2}u=c^2\left(u_{rr}+r^{-1}u_r\right).
\label{E13}
\end{equation}
We verify by direct differentiation (care is required) that
\begin{equation}
u(r,t)=\textstyle{{1\over2\pi c}}\delta(r-ct)
\label{E14}
\end{equation}
exactly solves (\ref{E13}). By superposing solutions of this form, we
construct a large class of solutions to (\ref{E3}) containing one arbitrary
function of two arguments:
\begin{equation}
u(x,y,t)=t\omega_{ct}[p],
\label{E15}
\end{equation}
where
\begin{equation}
\omega_{ct}[\phi]\equiv\frac{1}{2\pi}\int_0^{2\pi}d\theta\,\phi(x+ct\cos\theta,
y+ct\sin\theta).
\label{E16}
\end{equation}
Equation (\ref{E15}) is the exact solution to (\ref{E3}) for the initial
conditions (\ref{E2}) with $q(x,y)=u(x,y,0)=0$. Of course, (\ref{E15}) is not
the general solution to (\ref{E3}) because the general solution contains {\it
two} arbitrary functions of two arguments each. We cannot express the general
solution to (\ref{E3}) in Poisson form because (\ref{E13}) is time dependent,
and thus the time derivative of a solution is not a solution.

Huygens' principle holds for this special initial condition because $\omega_{ct}
[\phi]$ in (\ref{E16}) is determined by the values of $\phi(x_0,y_0)$ on the
{\it surface} but not in the {\it interior} of the 2-D light cone $(x-x_0)^2+
(y-y_0)^2=c^2t^2$. Indeed, (\ref{E14}) is the expanding-bubble wave created by a
flash bulb. There is no ringing effect if $q=0$. (This is the reverse of what
happens for the 1-D wave equation where Huygens' principle holds if $p=0$.)

{\bf Computational simulation of anomalous dispersion.}
The analytical arguments above can be verified by solving the equations
numerically. We have used the differential equation solver Comsol Multiphysics
to implement (\ref{E1}) for of three spatial dimensions [Fig.~\ref{F1}(a)], two
spatial dimensions [Fig.~\ref{F1}(b)], and for (\ref{E3}) [Fig.~\ref{F1}(c)]. A
delta-function localized initial condition used in the derivations above in
(\ref{E8}) and (\ref{E10})) cannot be modelled numerically. Instead, we use a
gaussian distribution with a spatial width, details of which are given in the
caption. Our simulations confirm the analytical results and show that solutions
to the wave equation (\ref{E1}) in 3-D obey Huygens' principle, while in 2-D
they exhibit anomalous dispersion. However, the modified 2-D wave equation
(\ref{E3}) also obeys Huygens' principle with no anomalous dispersion and no
trailing tail behind the pulses.

{\bf Wave equation with two space and two time dimensions.} The change of
variable $u(x,y,t)=tv(x,y,t)$ converts (\ref{E3}) to the wave equation
$$v_{tt}+t^{-1}v_t=c^2(v_{xx}+v_{yy}).$$
The left side of this equation is the two-dimensional radial {\it time
derivative} of the wave equation
\begin{equation}
v_{\alpha\alpha}+v_{\beta\beta}=c^2(v_{xx}+v_{yy}),
\label{E17}
\end{equation}
where ${\bf t}=(\alpha,\beta)$ is a {\it two-dimensional} vector time variable.
This shows that anomalous dispersion does not occur for solutions to a linear
homogeneous wave equation in a space of two time dimensions and two space
dimensions as long as the initial disturbance has a vanishing derivative.
Since the 3-D wave equation takes the form in (\ref{E17}) inside hyperbolic
metamaterials, these anisotropic media behave like a two-dimensional space with
no anomalous dispersion. Experimental observation of the absence of anomalous
dispersion in hyperbolic metamaterials is challanging because experiments to
date have used narrowband sources. (Any source that is narrowband will have
oscillations and the effect of this is to nearly cancel the $1/t$ tail that is
characteristic of anomalous dispersion in two-dimensional space.) However,
recent developments in ultrashort broadband pulses allow for generation of even
unipolar wave packets \cite{R10} which will be suitable for such experiments.)
Theoretical considerations developed here explain for the first time how
metamaterials can evade the anomalous dispersion intrinsic in 2-D
space.


\begin{thebibliography}{100}

\bibitem{R1} R.~Courant and D.~Hilbert, {\it Methods of Mathematical Physics},
Vol.~2 (Wiley, New York, 1962).

\bibitem{R2} P.~R.~Garabedian, {\it Partial Differential Equations}
(Wiley, New York, 1964).

\bibitem{R3} I.~I.~Smolyaninov and E.~E.~Narimanov, Phys. Rev. Lett.~{\bf 105},
067402 (2010).

\bibitem{R4} A. Poddubny, I. Iorsh, P. Belov, and Y. Kivshar,
Nat. Photonics {\bf 7}, 948 (2013).

\bibitem{R5} P. V. Kapitanova, P. Ginzburg, F. J. Rodr\'iguez-Fortu\~no,
D. S. Filonov, P. M. Voroshilov, P. A. Belov, A. N. Poddubny, Yu. S.
Kivshar, G. A. Wurtz, and A. V. Zayats, Nat. Commun. {\bf 5}, 3226 (2014).

\bibitem{R6} S. Ishii, A. V. Kildishev, E. Narimanov,
V. M. Shalaev, and V. P. Drachev, Laser Photon. Rev. {\bf 7}, 265 (2013).

\bibitem{R7} I. I. Smolyaninov, Y.-J. Hung, and E. E. Narimanov, Phys. Lett.
A {\bf 376}, 2575 (2012).

\bibitem{R8} I. I. Smolyaninov and Y.-J. Hung, Phys. Lett.
A {\bf 377}, 353 (2013).

\bibitem{R9} I. Bars and J. Terning, {\it Extra dimensions in space and
time} (Sringer, New York, 2010).

\bibitem{R10} R. M. Arkhipov, A. V. Pakhomov, I. V. Babushkin, M. V. Arkhipov,
Yu. A. Tolmachev, and N. N. Rosanov, J. Opt. Soc. Am. B {\bf 33}, 2518 (2016).

\end{thebibliography}
\end{document}